\documentclass{article}

\usepackage{arxiv}

\usepackage[utf8]{inputenc} 
\usepackage[T1]{fontenc}    
\usepackage{hyperref}       
\usepackage{url}            
\usepackage{booktabs}       
\usepackage{amsfonts}       
\usepackage{nicefrac}       
\usepackage{microtype}      
\usepackage{lipsum}
\usepackage{graphicx}
\usepackage{amsmath}
\graphicspath{ {./images/} }

\title{Proposal of a general framework to categorize continuous predictor variables}

\author{Irantzu Barrio \footnote{Corresponding author: {{E-mail: irantzu.barrio@ehu.eus}}, Tel.: +34-946012504, Address: Departamento de Matem\'aticas. Facultad de Ciencia y Tecnolog\'ia. Universidad del Pa\'is Vasco UPV/EHU. Barrio Sarriena s/n. 48940 Leioa.} $^{1,5}$\\
	Javier Roca-Pardi\~nas $^{2}$\\
	Cristobal Esteban$^{3}$\\    
	Maria Durban$^{4}$\\        
	\small{$^{1}$ Departamento de Matem\'aticas.}\\
	\small{Universidad del Pa\'is Vasco UPV/EHU}\\
	\small{$^{2}$ Departamento de Estad\'istica e Investigaci\'on Operativa.}\\
	\small{Universidade de Vigo}\\
	\small{$^{3}$ Servicio de Neumolog\'ia.}\\
	\small{Hospital Universitario de Galdakao}\\
	\small{$^{4}$ Departamento de Estad\'istica y Econometr\'ia}\\
	\small{Universidad Carlos III de Madrid}\\
	\small{$^{5}$  Basque Center for Applied Mathematics (BCAM)}
}

\usepackage[longnamesfirst]{natbib}
\bibpunct[, ]{(}{)}{;}{a}{,}{,}

\usepackage{color}
\usepackage{hyperref}
\usepackage{nicefrac}
\usepackage{lscape}  
\usepackage{multirow}
\usepackage{bigstrut}
\usepackage{caption}
\usepackage{subcaption}
\usepackage{longtable}  
\usepackage{booktabs}
\usepackage{multirow}
 \usepackage[table,xcdraw]{xcolor}
 \usepackage{longtable}

\renewcommand{\arraystretch}{1.3}

\begin{document}

\maketitle

\begin{abstract}
The use of discretized variables in the development of prediction models is a common practice, in part because the decision-making process is more natural when it is based on rules created from segmented models. Although this practice is perhaps more common in medicine, it is extensible to any area of knowledge where a predictive model helps in decision-making. Therefore, providing researchers with a useful and valid categorization method could be a relevant issue when developing prediction models. 

In this paper, we propose a new general methodology that can be applied to categorize a predictor variable in any regression model where the response variable belongs to the exponential family distribution. Furthermore, it can be applied in any multivariate context, allowing to categorize more than one continuous covariate simultaneously. In addition, a computationally very efficient method is proposed to obtain the optimal number of categories, based on a pseudo-BIC proposal. Several simulation studies have been conducted in which the efficiency of the method with respect to both the location and the number of estimated cut-off points is shown.

Finally, the categorization proposal has been applied to a real data set of 543 patients with chronic obstructive pulmonary disease from Galdakao Hospital's five outpatient respiratory clinics, who were followed up for 10 years. Exercise capacity is known to be an important predictor of adverse events in patients with COPD. In  this paper, we applied the proposed methodology to jointly categorize the continuous variables six-minute walking test and forced expiratory volume in one second in a multiple Poisson generalized additive model for the response variable rate of the number of hospital admissions by years of follow-up. The location and number of cut-off points  obtained were clinically validated as being in line with the categorizations used in the literature.

\end{abstract}


\section{Introduction}

Many clinical predictors are recorded as continuous variables, but in practice, clinical guidelines and scoring systems used to predict patients' morbidity and mortality often require the expression of these as categorical variables.  The development of clinical practice guidelines and scoring systems  is necessary in order to facilitate clinicians' decision-making process and help them reduce variability in clinical practice. 

In addition, in the development of prediction models for use in practice, continuous predictor variables are often categorized \citep{Barrio2020,spruit2012predicting}. These models provide estimates for individual risk of suffering some unfavorable event, which helps with decisions such as whether or not a given patient should be discharged, etc.  Therefore, how to categorize a continuous variable into a categorical one is of great interest to clinical researchers, clinicians, and healthcare professionals in general. 
	
	
When developing prediction models, the selection of the covariates (clinical variables) to be used in the model is essential, as well as good modeling of their relationship with the outcome. From a statistical point of view, categorizing continuous variables is not regarded as advisable, since it may entail a loss of information and power \citep{Altman2005,Royston2006}. Additionally, there are statistical modeling techniques such as generalized additive models (GAM) \citep{Hastie1986} which do not require any assumption of linearity between predictors and response variables, and so allow for the relationship between the predictor and the outcome to be modeled more appropriately. Nevertheless, one of the limitations that GAMs may have in practice is that it is not always easy to obtain decision rules based on this type of model, since the interpretability of GAMs may depend on the complexity of the risk functions \citep{hegselmann2020evaluation}. In this case, another possibility to achieve a more interpretable non-linear effect is to discretize the covariates, this is, turn it into a categorical covariate. In fact, based on the results of a survey of the epidemiological literature, in the $86\%$ of the papers included in the study, the primary continuous predictor was categorized, of which $78\%$ used 3 to 5 categories \citep{Turner2010}.

The estimation of a single cut-off point to dichotomize a continuous variable, a biomarker or a risk score, is a topic that has been discussed over a long period of time. \citet{miller1982} proposed to use the maximum chi-squared statistic (or minimum p-value) as the criterion to estimate the cut-off point for dichotomization purposes. \citet{altman1994} showed that the minimum p-value approach yields an increase in the type I error rate and thus proposed a correction for the minimum p-value formulae. Latter, \citet{faraggi1996} proposed a cross-validation approach to estimate the cut-off point based on the minimum p-value approach. On the other hand, \citet{hin1999} proposed the use of GAM and the curves obtained from them to dichotomize continuous variables in clinical prediction models. In parallel, proposals have been developed for the estimation of optimal cut-off points based on different criteria related to the predictive capacity of the predictive scores or variables to be categorized. For example, by maximizing the sensitivity or specificity parameters (i.e., probability of classifying correctly an individual with/without the event of interest), Youden or Kappa indexes \citep{youden1950,cohen1960, greiner2000}, among others. Some of these criteria have been implemented in the \texttt{OptimalCutpoints} \texttt{R} package \citep{Lopez2014}.

However, it may be preferable to use more than two categories, as this serves to reduce the loss of information and also allows the relationship between the covariate and the response variable to be better preserved.  In this regard, the selection of more than one cut-off point is a more recent topic of research. This has been often done based on percentiles, even though this is known to have drawbacks \citep{Bennette2012}. \cite{barrio2013} proposed a categorization methodology using GAM with P-spline smoothers, in which they considered creating at least one average-risk category along with high- and low-risk categories based on the GAM smooth function. However, the need for an additional category relied on a subjective decision. On the other hand, in the context where the response variable takes two possible values, \cite{tsuruta2006} proposed a parametric method for obtaining cut-off points based on the overall discrimination statistic, c-index \citep{Harrell1982}. The authors showed the optimal location of cut-off points in a case where the distribution of the continuous predictor variable is known, yet, in practice, continuous predictor variables do not usually respond to a known distribution. In the same context, \cite{Barrio2017} proposed a methodology to select the optimal cut-off points considering the maximal discrimination ability of the logistic regression model measured by the area under the receiver operative characteristic (ROC) curve (AUC), which for binomial response variables is equivalent to the c-index \citep{Bamber1975}. Unlike the \cite{tsuruta2006} proposal, the \cite{Barrio2017} proposal does not require distribution assumptions and can be used in any univariate or multiple modeling situation regardless of the distribution of the original continuous predictor. Furthermore, this methodology has also been extended for use in the Cox proportional hazards regression model \citep{Barrio2017a}, where different estimators for the concordance probability were considered in order to measure the discrimination ability of the categorical variable. Although this methodology allows selecting any possible number of cut-off points, either in a univariate or a multiple context, in many circumstances the number of categories in which to categorize the predictor variable is unclear. 

The optimal number of the cut-off points has been recently studied by \cite{chang2019} and \cite{Barrio2020}. The former considered the Cox proportional hazard regression model and proposed an AIC criterion to estimate the number of cut-off points, where the AIC values were corrected with cross-validation and Monte Carlo methods. On the other hand, the latter proposed a bootstrap-based hypothesis test in the logistic regression setting. In both cases, the proposed methodology has been developed for use in a specific regression model.

All the methods mentioned above have been developed to be applied in a specific context, be it logistic regression or the proportional hazards model for the most part. However, the response of interest may have a different distribution than the binomial and the interest need not be in modeling the time until the event occurs. Therefore, we have considered it necessary to develop a new methodology that allows categorizing a predictor variable in a more general context, in particular, where the response variable has a distribution belonging to the exponential family. In addition, the methodology proposed by \cite{Barrio2017}, although it can be applied in a multiple model, presents computational limitations when it comes to categorizing more than one variable at the same time, or the categorization of a variable according to the different factors taken by another variable with which there is a significant interaction. Thus, the methodology presented in this paper gives a solution to the limitations of the previous methods, allowing first the categorization of continuous variables in a different context than that of the logistic regression, and second, a computationally very efficient method when categorizing one or several continuous variables in any multiple contexts. This proposal is based on the idea that the smooth function of the GAM model can be approximated by a piecewise function, and provides a global framework to select both the location and the number of optimal cut-off points for any response variable with a distribution of the exponential family, which in addition, can be applied for any kind of multiple regression modeling approaches, including interactions or smooth functions. 

The rest of the paper is organized as follows. In Section \ref{notation} we introduce the main statistical models and notation that we will use in this document, starting with a description of the generalized linear and additive models (GLM, GAM).  In Section \ref{general_framework} we describe the proposal presented in this work, whose goal is twofold. On the one hand, to select the optimal location of the cut-off points and, on the other, to select the number of categories in which to categorize the continuous variable.  Section \ref{sim} outlines several simulation studies conducted with the aim of validating the proposed methodology as well as the criteria for selecting the number of cut-off points. We continue with the application to a real case study of patients with a chronic obstructive pulmonary disease  which is presented in Section \ref{secc_application}. Finally, the paper closes with a discussion in Section \ref{discussion} in which the main findings are reviewed and conclusions are drawn.

\section{Notation and preliminaries} \label{notation}
In this Section, we introduce the general notation  as well as the main statistical models we use throughout the paper.

The generalized linear model, (GLM), was introduced by  \cite{nelder1972} and further developed by \cite{McCullagh1989}. This approach has three main differences with respect to the linear model, which are: a) the normal distribution for the response variable is replaced by the exponential family distribution, b) a monotonic link function $g(\cdot)$ is used for modeling the relationship between the expected value of the response variable and the covariates and finally c) it uses an iterative reweighed least squares algorithm for the parameter's estimation.  The GLM model can be written as: 

\begin{equation}\label{eq1}
g(E(Y|{\bf X})) = \beta_0 +\beta_1 X_1 + \ldots + \beta_QX_Q,
\end{equation}
for $Y$ a response variable whose distribution belongs to the exponential family, $(X_1,\ldots,X_Q)$ a vector of  $Q$ covariates and  $g(\cdot)$ the link function. 

The GLM model assumes a linear relationship between the covariates $X_1, X_2, \ldots, X_Q$ and the link function $g(\cdot)$. However, in real life, effects are often not linear and we might be interested in a more flexible approach to model this relationship. This approach is called the generalized additive model (GAM), and was first introduced by \cite{Hastie1990}. The GAM model can be written as:

\begin{equation}\label{eq2}
g(E(Y|{\bf X})) = \beta_0 +f_1(X_1)  + \ldots + f_Q(X_Q),
\end{equation}
where $f_j(), \quad j=1,\ldots,Q$ stands for a non-parametric function applied to the continuous covariates (or some of). The idea is to let the data determine the relationship between the linear predictor $\eta=g(\mu(\bf X))$ and the explanatory variables rather than enforcing a linear relationship.

The original estimating method consisted of estimating the $f_j()$ by iterative smoothing of partial residuals with respect to the covariates. Later, the use of other methods such as kernel smoothers \citep{wand1994kernel} or spline-based smoothers \citep{green1993nonparametric} were proposed. Nevertheless, the penalized spline regression (P-splines, \cite{eilers1996}) has become one of the most popular methods to estimate the smooth functions $f_j()$ \citep{eilers2015}. The method is based on the representation of the smooth component as a combination of base functions and on the modification of the likelihood function by introducing a penalty based on differences between adjacent coefficients. In this context, the standard estimation method is the restricted maximum likelihood method (REML). However, in some circumstances, the penalties present an overlapping structure, with the same coefficients being penalized simultaneously by several smoothing parameters. In order to deal with this problem, \cite{rodriguez2019} have proposed an adaptive fast estimation method.

\begin{figure}
		\centering

			\includegraphics[height=0.25\textheight]{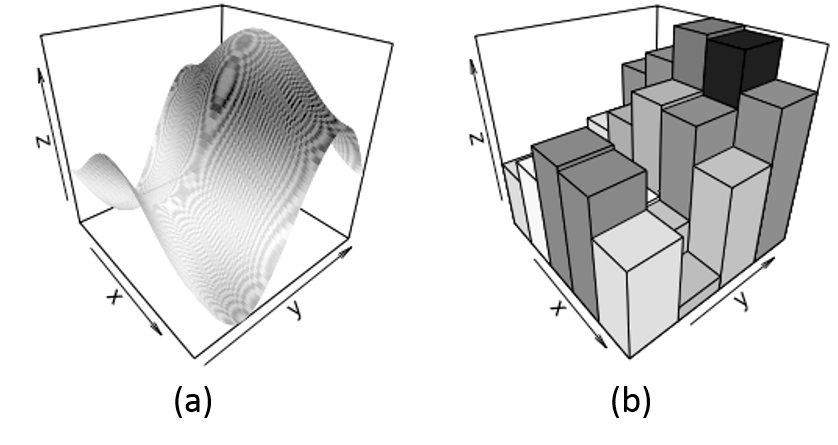}
			\caption{Illustrative example. a) Bivariate smooth relationship and b) Discretized relationship. }
			\label{fig:intro_disc_gam}
	\end{figure}

However, despite being a very flexible model, and being interpretable, in practice the GAM may not always be helpful when it comes to using the model during the decision-making process. In that case, it is easier if approximated by piecewise constants as shown in Figure \ref{fig:intro_disc_gam}. The present work is based on this idea.

\section{Proposed Methodology}\label{general_framework}

\subsection{Proposal to select the optimal location of cut-off points}\label{new_proposal}
Let $Y$ be a response variable belonging to the exponential family distribution, $X_1,\ldots,X_T$ a set of $T$ continuous predictor variables which we want to categorize (without loss of generality we consider the first $T$ in the set of $Q$ covariates) and  $(X_{T+1},\ldots,X_Q)$ a vector of $Q-T$ predictor variables which can be either continuous or categorical variables. For ease of notation, we will consider them hereinafter as continuous variables.

If it is considered a vector of $k_j$ cut-off points (ordered from lowest to highest) ${\bf c}_j=(c_{j,1},\ldots,c_{j,k_j})$ which defines the $k_j+1$ intervals for a covariate $X_j$ ($j=1,\ldots,T$) the effect $f_j(X_j)$ of the covariate $X_j$ can be approximated by

$$
f_j(X_j)\approx \beta_0+\sum_{s=1}^{k_j} \beta_{j,s} I \left \{ {c_{j,s} <X_j \leq c_{j,s+1}}\right\} \text{,} \quad \text{for} \quad j=1,\ldots,T
$$
where $\beta_{j,s}$, $j=1,\ldots, T$ and $s=1,\ldots,k_j$ are  unknown coefficients. Note that we have considered $[min(X_j),c_{j,1}]$ as the reference category, and, for simplicity of notation, we consider that $c_{j,k_j+1}=\infty$, $j =1,\ldots,T$. 

Therefore, the GAM model in \eqref{eq2} can be approximated by the model in  \eqref{piecewise}

\begin{equation}
g(E(Y|{\bf X, \bf c})) 	= \beta_0 +\sum_{j=1}^{T}\sum_{s=1}^{k_j} \beta_{j,s} I \left \{ {c_{j,s} <X_j \leq c_{j,s+1}}\right\}+   \sum_{j=T+1}^{Q} f_{j}(X_{j}).
		\label{piecewise}
\end{equation}

Under the model in equation \eqref{piecewise} it is very easy to interpret the effect of the covariates $X_j$ $j=1,\ldots, T$ on the response and makes it possible to establish rules in a much simpler way than the GAM model presented in equation  \eqref{eq2}. In addition, it facilitates the stratification of individuals, adjusting for the rest of covariables $X_j$ $j=T+1,\ldots,q$, in a simpler way than with a GAM. It also allows comparisons to be made with other studies where categorizations have been used.

The vector of cut-off points  ${\bf c}_j=(c_{j,1},\ldots,c_{j,k_j})$  in equation \eqref{piecewise}  can be established beforehand, or an estimation will have to be obtained from the data. Note here that sometimes ${\bf c}_j$ may not be known but the value of $k_j$ may be established a priori. In other cases both ${\bf c}_j$ and $k_j$ may be unknown, for $j=1,\ldots,T$. We describe below our proposal for estimating both the location and the number of cutoff points when both are unknown.

\begin{description}
\setlength\itemsep{2em}
\item[\textbf{Step 1}] Given a sample  ${\{({\bf x}_i,y_i)\}}_{i=1}^n$, we propose to estimate the GAM model in equation \eqref{eq2} using p-splines and an adaptive fast estimation algorithm \citep{rodriguez2019} to estimate the smooth functions $f_j(), \quad j=1,\ldots,Q$.

\begin{equation}\label{eq4}
\widehat{g(E(Y|{\bf X}))} = \hat{\beta_0} +\hat{f}_1(X_1)  + \ldots +\hat{f}_T(X_T) + \hat{f}_{T+1}(X_{T+1}) + \ldots + \hat{f}_Q(X_Q).
\end{equation}

\item[For $j=1,...,T$:]
			
\item[\textbf{Step 2}] Obtain the estimates $\hat f_j (x_{j1}), \ldots, \hat f_j (x_{jn})$ and the standard errors $\hat \sigma_j (x_{j1}), \ldots, \hat \sigma_j (x_{jn})$, resulting from the fit of the model in \eqref{eq4}.

\item[\textbf{Step 3}]  Given ${\bf c}_j$ we define the \textbf{weighted mean squared error} 
			
			\[
			WMSE ({\bf c}_j)=\sum_{i=1}^n \frac {\left({\hat f_j(x_{ji})-\bar{f_j}^{{\bf c}_j} (x_{ji}) }\right)^2} {\hat{\sigma_j}^2 (x_{ji})}
			\]
			where
			\[
			\bar{f_j}^{{\bf c}_j} (x_{ji})= \sum_{s=0}^{k_j} I \left\{ c_{j,s} < x_{ji} \leq c_{j,s+1}\right\}  \left[  \frac{ \sum_{l = 1}^n   \hat{f_j}(x_{jl})  I \left \{ c_{j,s} <x_{jl} \leq c_{j,s+1}\right \} } {\sum_{l=1}^n   I \left \{ c_{j,s} <x_{jl} \leq c_{j,s+1}\right\} } \right]			
			\]
			is the mean of the estimated $\hat f_j$ in the category $R^j_s=(c_{j,s},c_{j,s+1}]$ for $s=0,\ldots,k_j$.

\item[\textbf{Step 4}]  Estimate the vector of cut-off points ${\bf c}_j$ 
	
   \noindent
	\textbf{Initialize}: Compute the initial estimates $(\hat c_{j,1}^{0}, \ldots,\hat c_{j,k_j}^{0})$.
			
			\noindent
			\textbf{Step 4.1}: Cycle $r=1,\ldots,k_j$ calculating the update
			
			\[
			\hat c_{j,r}= \text{argmin}_c WMSE(\hat c_{j,1}, \ldots, \hat c_{j,r-1} , c , 
			\hat c_{j,r+1}^0,\ldots,\hat c_{j,k_j}^0)
			\]
			
			\noindent
			\textbf{Step 4.2}: Repeat \textbf{Step 4.1} replacing $(\hat c_{j,1}^{0}, \ldots,\hat c_{j,k_j}^{0})$ by $(\hat c_{j,1}, ...,\hat c_{j,k_j})$ until the convergence is obtained.
\end{description}

\subsection{Proposal to select the number of categories}
Bayesian information criterion (BIC) is a useful solution to the problem posed by direct comparison of likelihoods when selecting a model (since the likelihood will never decrease even if redundant parameters are included in the model). It was originally proposed by \cite{schwarz1978} as a modification of Akaike's information criterion (AIC), by introducing a larger penalty term for the  effective dimension of the model. 

For any given model $\Theta$, the BIC is defined as follows:

\begin{equation}
BIC=\phi log(n) - 2log(L(\Theta))
\end{equation}
where $\phi$ and $L(\Theta)$ stand for the effective dimension and likelihood of model $\Theta$, respectively. Note that in the presence of estimated smooth functions $f_j$ the effective dimension accounts for penalization, using the correction proposed by \cite{wood2016smoothing}.

Considering $\hat{{\bf c}_j}=(\hat{c}_{j,1},\ldots,\hat{c}_{j,k_j}) $ the vector of the optimal cut-off points for the covariate $X_j$ $j=1,\ldots,k_j$ estimated on the basis of the method presented in Section \ref{new_proposal} above (and a sample  ${\{({\bf x}_i, y_i)\}}_{i=1}^n$), the estimated model for $X_j$ $j=1,\ldots, T$ categorized based on these cut-off points would be,

\begin{equation}
\widehat{g(E(Y|{\bf x, \bf c})}) 	= \hat{\beta}_0 +\sum_{j=1}^{T}\sum_{s=1}^{k_j} \hat{\beta}_{j,s} I \left \{ {c_{j,s} <x_j \leq c_{j,s+1}}\right\}+   \sum_{j=T+1}^{Q} \hat{f}_{j}(x_{j}).
		\label{modelo_cat}
\end{equation}

Thus, the BIC for the model in equation \eqref{modelo_cat} would be

\begin{equation}
BIC^{{\bf c}}=\phi log(n) - 2log(L(\bf c)),
\end{equation}
with $L(\bf c)$ and $\phi$, the likelihood and effective dimension of the model represented in equation \eqref{modelo_cat}, respectively.

However, since in addition to model parameters ($\beta_{j,s}$ $j=1,\ldots,T$, $s=1,\ldots, k_j$) and smooth functions ($f_1,\ldots,f_Q$), also the cut-off points ($\hat{{\bf c}_j}=(\hat{c}_{j,1},\ldots,\hat{c}_{j,k_j}) $ $j=1,\ldots,T$)  have been estimated,  we define a pseudo-BIC, which considers the estimated number of cut-off points

\begin{equation}\label{pseudo_BIC}
BIC_{pseudo}^{{\bf c}}=BIC^{{\bf c}} + log(n) \left(  \sum_{j=1}^T k_j \right)  .
\end{equation}


We propose to use the minimum $BIC_{pseudo}$ as the criteria to select the number of cut-off points for each variable $X_j$, $j=1,...,T$, by penalizing the effect of introducing an extra category (i.e, cut-off point) for the categorized variable. Let's denote ${\bf nc}=(nc_1, \ldots, nc_T)$ as the vector of the number of cut-off points needed to categorize the predictor variables $X_1,\ldots,X_T$, then:

\[
{\bf nc} = argmin BIC_{pseudo}^{{\bf c}}. 
\] 

\section{Simulation study}\label{sim}

In this section, we present a simulation study with two different goals: 1) performance evaluation with regard to location, and 2) performance evaluation with regard to the selected number of cut-off points. Therefore, we conducted a simulation study considering a theoretically defined piece-wise multiple model for a binomial response variable, considering different theoretical numbers of cut-off points for two continuous covariates we aimed to categorize.

The simulation study was performed in (64 bit) R 4.1.2 and run on a workstation equipped with 32 GB of RAM, Intel Core i7-8700 processor (3.20 GHz), and the Windows 10 operating system.

\textbf{Scenarios and set-up}

Let, $\boldsymbol{X}=(X_1,X_2,X_3)$ be a vector of continuous covariates, and $Y \sim Bin(1, p(\boldsymbol{X}))$, given that 

$$
log \left( \frac{p(\boldsymbol{X})}{1- p(\boldsymbol{X})} \right)= f_1(X_1) +f_2(X_2) + 0.1X_3.
$$

The continuous covariates $X_1$ and $X_2$ were drawn from a uniform distribution in the interval $[-2,2]$, while $X_3$ was drawn from a uniform distribution in the interval $[-1,1]$. The  $f_j(X_j)$  was defined as a piece-wise constant function for $j=1,2$ such that:

	\[
	f_j(X_j)=\left \{ {
		\begin{tabular}{lcl}
		$a_{j,1}$ & if & $X_j \leq c_{j,1}$\\
		$a_{j,2}$  & if & $c_{j,1}< X_j \leq c_{j,2}$\\
		\vdots & \\
		$a_{j,k_j}$  & if & $c_{j,k_j-1}< X_j \leq c_{j,k_j}$\\
		$a_{j,k_j+1}$ & if & $X_j>c_{j,k_j}$\\
		\end{tabular}
	}\right. 
	\]
	\noindent
 The vector of the theoretical cut-off points  $\boldsymbol{c}_j=(c_{j,1},\ldots, c_{j,k_j})$ (for $j=1,2$) was defined to obtain an equidistant sequence of values in the interval $[-2,2]$, leaving aside the extreme values.
	
Four different scenarios were considered based on different theoretical numbers of cut-off points $k_j$ for $X_j$, $j=1,2$, which are summarized in Table \ref{tab:scenarios}.

\begin{table}[h]
\caption{Description of the scenarios considered. From left to right, (a) theoretical number of cut-off points considered in each scenario, $k_1$ and $k_2$ for $X_1$ and $X_2$, respectively; and (b) parametrization considered for $f_1(X_1)$ and $f_2(X_2)$.}
\label{tab:scenarios}

\begin{subtable}[h]{0.25\textwidth}
\centering
\begin{tabular}{lcc}
\hline
\textbf{Scenario}     & $k_1$ & $k_2$ \\
 & & \\ \hline
\textbf{S1}   & 2     & 1     \\
\textbf{S2}  & 2     & 2     \\
\textbf{S3} & 3     & 1     \\
\textbf{S4}  & 3     & 2     \\ \bottomrule
\end{tabular}
\caption{}
\label{tab:scenariosA}
\end{subtable}  \hfill
\begin{subtable}[h]{0.65\textwidth}
\centering
\begin{tabular}{rr|rr|rr|rr} \hline
\multicolumn{2}{c|}{$k_1=2$}                                                     & \multicolumn{2}{c|}{$k_1=3$}                                                     & \multicolumn{2}{c|}{$k_2=1$}                                                     & \multicolumn{2}{c}{$k_2=2$}                                                     \\
\multicolumn{1}{c}{$\boldsymbol{a}_1$} & \multicolumn{1}{c|}{$\boldsymbol{c}_1$} & \multicolumn{1}{c}{$\boldsymbol{a}_1$} & \multicolumn{1}{c|}{$\boldsymbol{c}_1$} & \multicolumn{1}{c}{$\boldsymbol{a}_2$} & \multicolumn{1}{c|}{$\boldsymbol{c}_2$} & \multicolumn{1}{c}{$\boldsymbol{a}_2$} & \multicolumn{1}{c}{$\boldsymbol{c}_2$} \\ \hline
1.5  & $-0.67$      & 1.5    & $-1$ & $-2$ & 0  & $-2$               & $-0.67$          \\
0       & 0.67   & 0   & 0  & 0   &        & 0  & 0.67          \\
1.5   &    & 1.5    & 1          &   &  & 2  &                   \\
 &    & 3   &      & &   &      &                                        \\ \bottomrule
\end{tabular}
\caption{}
\label{tab:scenariosB}
\end{subtable}
\end{table}

For sample sizes $n \in \left\{(500,1000,2000) \right\}$ and the four scenarios considered $m \in \left\{(S1,S2,S3,S4) \right\}$, $R=500$ independent samples  $\left\{({\bf x}_i, y_i^{m}) \right\}_{i=1}^n$  were generated with $y_i^{m}=Bernoulli(P(\boldsymbol{X}=\boldsymbol{x}_i))$.  The performance of the proposed methodology to estimate the location of the cut-off points was evaluated by means of the bias and mean square error (MSE) of the estimated optimal cut-off points for each iteration as follows:

\[
MSE_{r}=\frac{1}{k_1+k_2}\sum_{j=1}^{2}\sum_{s=1}^{k_j}(\hat{c}^r_{j,s}-c_{j,s})^2, \quad \text{for} \quad r=1,\ldots, R.
\]

where $\hat c^r_{j,s}$ is the estimated $s^{th}$ optimal cut-off point in the simulation $r$, and $c_{j,s}$ is the $s^{th}$ theoretical cut-off point,  for the covariate $X_j$ with $j=1,2$.

The number of cut-off points was selected considering those for which the $BIC_{pseudo}$ was minimized, as long as the consecutive categories were statistically significant with $\alpha=0.05$. If this was not met, this combination of cut-off points was not considered.

\textbf{Results}

\begin{landscape}
    	\begin{figure}
		\centering

			\includegraphics[height=0.8\textheight]{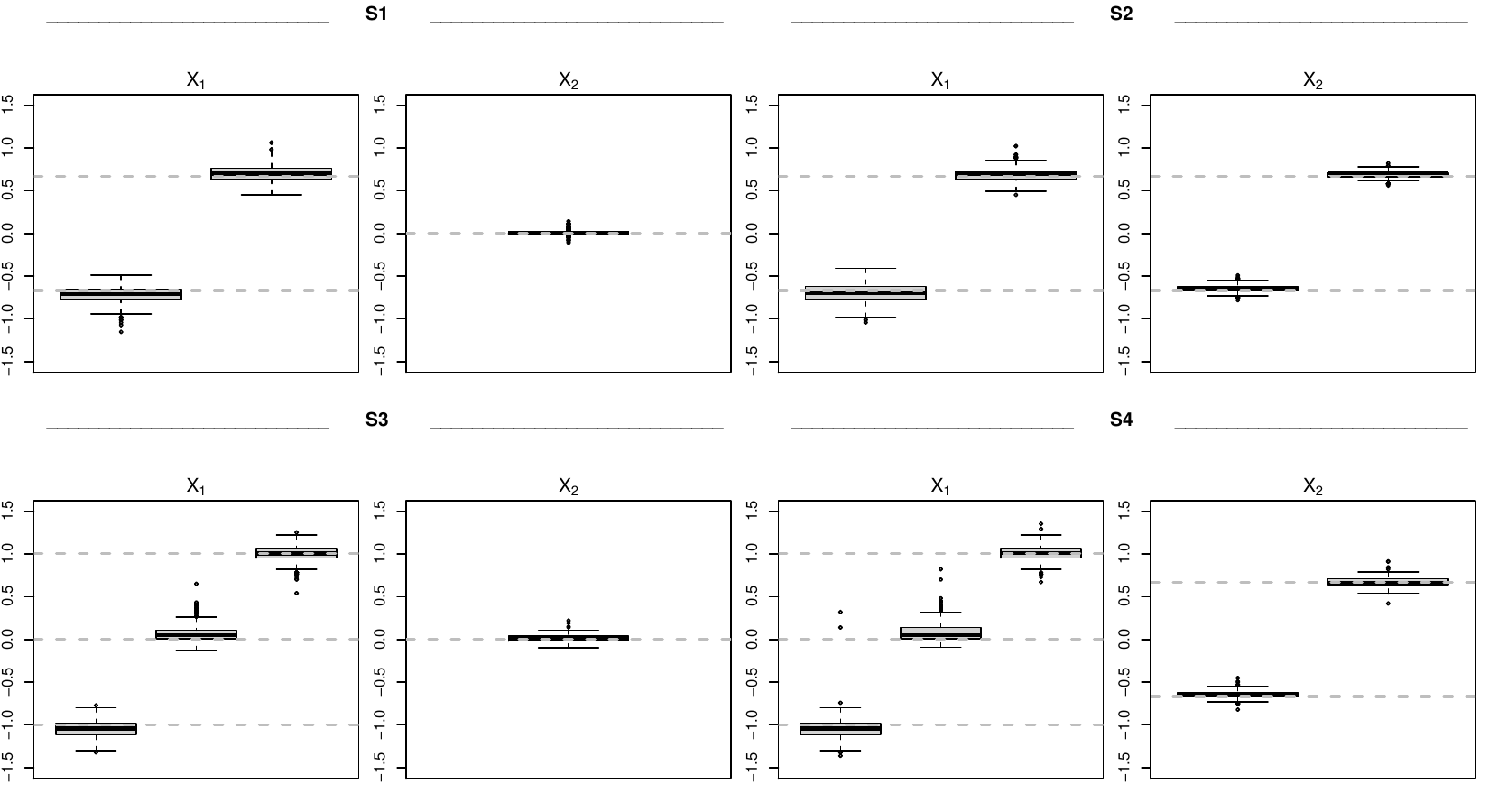}
			\caption{Boxplots of the estimated cut-off points for a sample size of $n=1000$ in scenarios S1,S2,S3 and S4.}
			\label{fig:location}
	\end{figure}	
\end{landscape}

The numerical results obtained for the estimated optimal cut-off points for different sample sizes ($n=500$, $n=1000$ and $n=2000$), scenarios ($ m \in \left\{(S1,S2,S3,S4) \right\}$), and $R=500$ replicates are summarized in Table \ref{tab:result_location}. In addition, Figure \ref{fig:location} depicts the boxplot of the estimated optimal cut-off points over 500 simulated data sets for a sample size of $n=1000$ and all the scenarios considered.


\newpage
\renewcommand{\arraystretch}{1.1}
\begin{longtable}{cccrrrrrr}
\caption{Numerical results for the location of the estimated optimal cut-off points for scenarios $ m \in \left\{S1,S2,S3,S4 \right\}$ and sample sizes $ n \in \left\{500,1000,2000 \right\}$.}
\label{tab:result_location} \\
\hline 
\multirow{2}{*}{\textbf{Scenario}} & \multirow{2}{*}{\textbf{sample size}} & \multirow{2}{*}{\textbf{Variable}} & \multicolumn{1}{c}{\multirow{2}{*}{\textbf{truec}}} & \multicolumn{3}{c}{\textbf{Estimated cut-off   point}}                                                  & \multicolumn{2}{c}{\textbf{MSE}}                \\ \cline{5-9} 
                                   &                                       &                                    & \multicolumn{1}{c}{}                                & \multicolumn{1}{c}{\textbf{Mean}} & \multicolumn{1}{c}{\textbf{Sd}} & \multicolumn{1}{c}{\textbf{Bias}} & \textbf{Mean}          & \textbf{Sd}            \\ \hline \endhead
\multirow{9}{*}{S1}                & \multirow{3}{*}{500}                  & \multirow{2}{*}{$X_1$}             & -0.667                                              & -0.804                            & 0.173                           & -0.137                            & \multirow{3}{*}{0.027} & \multirow{3}{*}{0.025} \\
                                   &                                       &                                    & 0.667                                               & 0.750                             & 0.139                           & 0.083                             &                        &                        \\
                                   &                                       & $X_2$                              & 0.000                                               & 0.001                             & 0.075                           & 0.001                             &                        &                        \\ \cline{2-9} 
                                   & \multirow{3}{*}{1000}                 & \multirow{2}{*}{$X_1$}             & -0.667                                              & -0.720                            & 0.102                           & -0.053                            & \multirow{3}{*}{0.008} & \multirow{3}{*}{0.011} \\
                                   &                                       &                                    & 0.667                                               & 0.696                             & 0.085                           & 0.029                             &                        &                        \\
                                   &                                       & $X_2$                              & 0.000                                               & 0.010                             & 0.033                           & 0.010                             &                        &                        \\ \cline{2-9} 
                                   & \multirow{3}{*}{2000}                 & \multirow{2}{*}{$X_1$}             & -0.667                                              & -0.666                            & 0.044                           & 0.001                             & \multirow{3}{*}{0.001} & \multirow{3}{*}{0.001} \\
                                   &                                       &                                    & 0.667                                               & 0.676                             & 0.045                           & 0.009                             &                        &                        \\
                                   &                                       & $X_2$                              & 0.000                                               & 0.009                             & 0.016                           & 0.009                             &                        &                        \\ \hline
\multirow{12}{*}{S2}               & \multirow{4}{*}{500}                  & \multirow{2}{*}{$X_1$}             & -0.667                                              & -0.804                            & 0.177                           & -0.138                            & \multirow{4}{*}{0.027} & \multirow{4}{*}{0.049} \\
                                   &                                       &                                    & 0.667                                               & 0.723                             & 0.147                           & 0.057                             &                        &                        \\
                                   &                                       & \multirow{2}{*}{$X_2$}             & -0.667                                              & -0.708                            & 0.129                           & -0.042                            &                        &                        \\
                                   &                                       &                                    & 0.667                                               & 0.723                             & 0.102                           & 0.056                             &                        &                        \\ \cline{2-9} 
                                   & \multirow{4}{*}{1000}                 & \multirow{2}{*}{$X_1$}             & -0.667                                              & -0.715                            & 0.114                           & -0.048                            & \multirow{4}{*}{0.007} & \multirow{4}{*}{0.008} \\
                                   &                                       &                                    & 0.667                                               & 0.694                             & 0.081                           & 0.027                             &                        &                        \\
                                   &                                       & \multirow{2}{*}{$X_2$}             & -0.667                                              & -0.642                            & 0.046                           & 0.025                             &                        &                        \\
                                   &                                       &                                    & 0.667                                               & 0.685                             & 0.045                           & 0.018                             &                        &                        \\ \cline{2-9} 
                                   & \multirow{4}{*}{2000}                 & \multirow{2}{*}{$X_1$}             & -0.667                                              & -0.671                            & 0.054                           & -0.004                            & \multirow{4}{*}{0.002} & \multirow{4}{*}{0.002} \\
                                   &                                       &                                    & 0.667                                               & 0.676                             & 0.050                           & 0.009                             &                        &                        \\
                                   &                                       & \multirow{2}{*}{$X_2$}             & -0.667                                              & -0.661                            & 0.028                           & 0.006                             &                        &                        \\
                                   &                                       &                                    & 0.667                                               & 0.678                             & 0.028                           & 0.012                             &                        &                        \\ \hline
\multirow{12}{*}{S3}               & \multirow{4}{*}{500}                  & \multirow{3}{*}{$X_1$}             & -1.000                                              & -1.077                            & 0.191                           & -0.077                            & \multirow{4}{*}{0.029} & \multirow{4}{*}{0.068} \\
                                   &                                       &                                    & 0.000                                               & 0.161                             & 0.151                           & 0.161                             &                        &                        \\
                                   &                                       &                                    & 1.000                                               & 0.959                             & 0.131                           & -0.041                            &                        &                        \\
                                   &                                       & $X_2$                              & 0.000                                               & 0.004                             & 0.084                           & 0.004                             &                        &                        \\ \cline{2-9} 
                                   & \multirow{4}{*}{1000}                 & \multirow{3}{*}{$X_1$}             & -1.000                                              & -1.052                            & 0.089                           & -0.052                            & \multirow{4}{*}{0.010} & \multirow{4}{*}{0.012} \\
                                   &                                       &                                    & 0.000                                               & 0.085                             & 0.102                           & 0.085                             &                        &                        \\
                                   &                                       &                                    & 1.000                                               & 0.989                             & 0.093                           & -0.011                            &                        &                        \\
                                   &                                       & $X_2$                              & 0.000                                               & 0.013                             & 0.041                           & 0.013                             &                        &                        \\ \cline{2-9} 
                                   & \multirow{4}{*}{2000}                 & \multirow{3}{*}{$X_1$}             & -1.000                                              & -1.004                            & 0.044                           & -0.004                            & \multirow{4}{*}{0.002} & \multirow{4}{*}{0.003} \\
                                   &                                       &                                    & 0.000                                               & 0.040                             & 0.050                           & 0.040                             &                        &                        \\
                                   &                                       &                                    & 1.000                                               & 1.000                             & 0.050                           & 0.000                             &                        &                        \\
                                   &                                       & $X_2$                              & 0.000                                               & 0.010                             & 0.018                           & 0.010                             &                        &                        \\ \hline
\multirow{10}{*}{S4}               & \multirow{5}{*}{500}                  & \multirow{3}{*}{$X_1$}             & -1.000                                              & -1.046                            & 0.252                           & -0.046                            & \multirow{5}{*}{0.034} & \multirow{5}{*}{0.075} \\
                                   &                                       &                                    & 0.000                                               & 0.166                             & 0.169                           & 0.166                             &                        &                        \\
                                   &                                       &                                    & 1.000                                               & 0.966                             & 0.129                           & -0.034                            &                        &                        \\
                                   &                                       & \multirow{2}{*}{$X_2$}             & -0.667                                              & -0.717                            & 0.120                           & -0.050                            &                        &                        \\
                                   &                                       &                                    & 0.667                                               & 0.724                             & 0.093                           & 0.057                             &                        &                        \\ \cline{2-9} 
                                   & \multirow{5}{*}{1000}                 & \multirow{3}{*}{$X_1$}             & -1.000                                              & -1.046                            & 0.123                           & -0.046                            & \multirow{5}{*}{0.010} & \multirow{5}{*}{0.029} \\
                                   &                                       &                                    & 0.000                                               & 0.089                             & 0.112                           & 0.089                             &                        &                        \\
                                   &                                       &                                    & 1.000                                               & 1.000                             & 0.087                           & 0.000                             &                        &                        \\
                                   &                                       & \multirow{2}{*}{$X_2$}             & -0.667                                              & -0.638                            & 0.046                           & 0.029                             &                        &                        \\
                                   &                                       &                                    & 0.667                                               & 0.679                             & 0.052                           & 0.012                             &                        &                        \\ \cline{2-9} 
\multirow{5}{*}{S4}                  & \multirow{5}{*}{2000}                 & \multirow{3}{*}{$X_1$}             & -1.000                                              & -1.012                            & 0.048                           & -0.012                            & \multirow{5}{*}{0.002} & \multirow{5}{*}{0.002} \\
                                   &                                       &                                    & 0.000                                               & 0.033                             & 0.053                           & 0.033                             &                        &                        \\
                                   &                                       &                                    & 1.000                                               & 1.001                             & 0.045                           & 0.001                             &                        &                        \\
                                   &                                       & \multirow{2}{*}{$X_2$}             & -0.667                                              & -0.652                            & 0.033                           & 0.015                             &                        &                        \\
                                   &                                       &                                    & 0.667                                               & 0.675                             & 0.031                           & 0.008                             &                        &                        \\
\hline
\end{longtable}

 Simulation results suggest that, in general, the proposed methodology performs satisfactorily in regard to the estimation of the location of the cut-off points. As could be expected, as the sample size increases the bias and the MSE decrease. Nonetheless, the increase in the number of theoretical cut-off points does not significantly increase the MSE. For instance, in Scenario S1 (where $k_1=2$ and $k_2=1$), mean MSE values of 0.027 and 0.001 have been obtained for sample sizes of 500 and 2000, respectively. On the other hand, in scenario S2  (where $k_1=2$ and $k_2=2$) mean MSE values of 0.027 and 0.002 have been obtained for sample sizes of 500 and 2000, respectively. Biases above 0.1 were only obtained in samples of size 500, and in all cases, only at one of the cut-off points.

Regarding the number of cut-off points selected, Table \ref{tab:pseudoBIC} provides the results obtained where the percentage of simulations (out of 500) in which each of the possible number (combination) of cut-off points has been selected. As can be seen, for sample sizes of 1000 and 2000, the true number of cut-off points are selected in more than 90$\%$ of simulations. This percentage decreases as the theoretical number of cut-off points increases. For example, going from scenario S1 to S2 (where the covariate $X_2$ has 2 theoretical cut-off points instead of 1) for a sample size of 2000, we go from a 96$\%$ success rate to $91.2\%$. On the other hand, when the sample size is 500, different results have been obtained depending on the scenario. In scenarios S1 and S2, the number of theoretical cut-off points was obtained in 87.4$\%$ and 82.2$\%$ of the replicates, respectively. However, in scenarios S3 and S4, the method tends to select fewer cut-off points (categories) for the variable $X_1$, which has been theoretically defined with 3 theoretical cut-off points. Note here that those cases in which an additional category is not statistically significant have not been considered, and therefore, we understand that with small samples the model tends to be fitted with a smaller number of categories. 


\begin{landscape}

\begin{table}[]
			\caption{Selected number of points based on the Pseudo-BIC criteria for sample sizes $ n \in \left\{(500,1000,2000) \right\}$ and scenarios $ m \in \left\{(S1,S2,S3,S4) \right\}$.}
			\label{tab:pseudoBIC}
\begin{tabular}{@{}cllllllllllllllllllll@{}}
\toprule
\multicolumn{1}{l}{}            &                                                & \multicolumn{4}{c}{\textbf{Scenario I}}                                                                &  & \multicolumn{4}{c}{\textbf{Scenario II}}                                                               &  & \multicolumn{4}{c}{\textbf{Scenario III}}                                                              &  & \multicolumn{4}{c}{\textbf{Scenario IV}}                                                               \\ \midrule
\multicolumn{1}{l}{}            & nc                                             & \multicolumn{4}{c}{$X_2$}                                                                              &  & \multicolumn{4}{c}{$X_2$}                                                                              &  & \multicolumn{4}{c}{$X_2$}                                                                              &  & \multicolumn{4}{c}{$X_2$}                                                                              \\
\multicolumn{1}{l}{Sample Size} & \multicolumn{1}{c}{$X_1$}                      & 1                                                            & 2                           & 3   & 4   &  & 1   & 2                                                            & 3                           & 4   &  & 1                                                            & 2                           & 3   & 4   &  & 1   & 2                                                            & 3                           & 4   \\ \cmidrule(lr){3-6} \cmidrule(lr){8-11} \cmidrule(lr){13-16} \cmidrule(l){18-21} 
$n =500$                        & \multicolumn{1}{l|}{\cellcolor[HTML]{FFFFFF}1} & \cellcolor[HTML]{EFEFEF}9.0                                  & \cellcolor[HTML]{FFFFFF}0.0 & 0.0 & 0.0 &  & 0.2 & \cellcolor[HTML]{EFEFEF}11.0                                 & 0.6                         & 0.0 &  & \cellcolor[HTML]{EFEFEF}24.2                                 & 0.2                         & 0.0 & 0.0 &  & 0.0 & \cellcolor[HTML]{EFEFEF}26.2                                 & 1.0                         & 0.0 \\
                                & \multicolumn{1}{l|}{\cellcolor[HTML]{FFFFFF}2} & \cellcolor[HTML]{9B9B9B}{\color[HTML]{FFFFFF} \textbf{87.4}} & \cellcolor[HTML]{FFFFFF}1.6 & 0.2 & 0.0 &  & 0.0 & \cellcolor[HTML]{9B9B9B}{\color[HTML]{FFFFFF} \textbf{82.2}} & \cellcolor[HTML]{EFEFEF}5.0 & 0.0 &  & \cellcolor[HTML]{EFEFEF}13.8                                 & 0.4                         & 0.4 & 0.0 &  & 0.0 & \cellcolor[HTML]{EFEFEF}14.2                                 & 0.4                         & 0.0 \\
                                & \multicolumn{1}{l|}{\cellcolor[HTML]{FFFFFF}3} & \cellcolor[HTML]{FFFFFF}1.6                                  & \cellcolor[HTML]{FFFFFF}0.2 & 0.0 & 0.0 &  & 0.0 & 0.8                                                          & 0.2                         & 0.0 &  & \cellcolor[HTML]{9B9B9B}{\color[HTML]{FFFFFF} \textbf{58.0}} & \cellcolor[HTML]{EFEFEF}2.8 & 0.0 & 0.0 &  & 0.0 & \cellcolor[HTML]{9B9B9B}{\color[HTML]{FFFFFF} \textbf{55.2}} & \cellcolor[HTML]{EFEFEF}2.2 & 0.0 \\
                                & \multicolumn{1}{l|}{\cellcolor[HTML]{FFFFFF}4} & 0.0                                                          & 0.0                         & 0.0 & 0.0 &  & 0.0 & 0.0                                                          & 0.0                         & 0.0 &  & 0.2                                                          & 0.0                         & 0.0 & 0.0 &  & 0.0 & 0.8                                                          & 0.0                         & 0.0 \\
\multicolumn{1}{l}{}            &                                                &                                                              &                             &     &     &  &     &                                                              &                             &     &  &                                                              &                             &     &     &  &     &                                                              &                             &     \\
$n =1000$                       & \multicolumn{1}{l|}{1}                         & \cellcolor[HTML]{FFFFFF}0.0                                  & \cellcolor[HTML]{FFFFFF}0.0 & 0.0 & 0.0 &  & 0.0 & 0.0                                                          & 0.0                         & 0.0 &  & 1.2                                                          & 0.0                         & 0.0 & 0.0 &  & 0.0 & 1.6                                                          & 0.0                         & 0.0 \\
                                & \multicolumn{1}{l|}{2}                         & \cellcolor[HTML]{9B9B9B}{\color[HTML]{FFFFFF} \textbf{95.0}} & \cellcolor[HTML]{FFFFFF}1.6 & 0.0 & 0.0 &  & 0.0 & \cellcolor[HTML]{9B9B9B}{\color[HTML]{FFFFFF} \textbf{94.0}} & \cellcolor[HTML]{EFEFEF}2.6 & 0.2 &  & 1.2                                                          & 0.0                         & 0.0 & 0.0 &  & 0.0 & \cellcolor[HTML]{EFEFEF}2.0                                  & 0.0                         & 0.0 \\
                                & \multicolumn{1}{l|}{3}                         & \cellcolor[HTML]{EFEFEF}2.8                                  & \cellcolor[HTML]{FFFFFF}0.4 & 0.0 & 0.0 &  & 0.0 & \cellcolor[HTML]{EFEFEF}3.0                                  & 0.2                         & 0.0 &  & \cellcolor[HTML]{9B9B9B}{\color[HTML]{FFFFFF} \textbf{92.4}} & \cellcolor[HTML]{EFEFEF}2.6 & 0.2 & 0.0 &  & 0.0 & \cellcolor[HTML]{9B9B9B}{\color[HTML]{FFFFFF} \textbf{90.2}} & \cellcolor[HTML]{EFEFEF}2.0 & 0.0 \\
                                & \multicolumn{1}{l|}{4}                         & 0.2                                                          & 0.0                         & 0.0 & 0.0 &  & 0.0 & 0.0                                                          & 0.0                         & 0.0 &  & \cellcolor[HTML]{EFEFEF}2.0                                  & 0.4                         & 0.0 & 0.0 &  & 0.0 & \cellcolor[HTML]{EFEFEF}4.2                                  & 0.0                         & 0.0 \\
\multicolumn{1}{l}{}            &                                                &                                                              &                             &     &     &  &     &                                                              &                             &     &  &                                                              &                             &     &     &  &     &                                                              &                             &     \\
$n =2000$                       & \multicolumn{1}{l|}{1}                         & 0.0                                                          & 0.0                         & 0.0 & 0.0 &  & 0.0 & 0.0                                                          & 0.0                         & 0.0 &  & 0.0                                                          & 0.0                         & 0.0 & 0.0 &  & 0.0 & 0.0                                                          & 0.0                         & 0.0 \\
                                & \multicolumn{1}{l|}{2}                         & \cellcolor[HTML]{9B9B9B}{\color[HTML]{FFFFFF} \textbf{96.0}} & 1.2                         & 0.0 & 0.0 &  & 0.0 & \cellcolor[HTML]{9B9B9B}{\color[HTML]{FFFFFF} \textbf{91.2}} & 3.8                         & 0.2 &  & 0.0                                                          & 0.0                         & 0.0 & 0.0 &  & 0.0 & 0.0                                                          & 0.0                         & 0.0 \\
                                & \multicolumn{1}{l|}{3}                         & 2.8                                                          & 0.0                         & 0.0 & 0.0 &  & 0.0 & 4.8                                                          & 0.0                         & 0.0 &  & \cellcolor[HTML]{9B9B9B}{\color[HTML]{FFFFFF} \textbf{93.4}} & 1.6                         & 0.0 & 0.0 &  & 0.0 & \cellcolor[HTML]{9B9B9B}{\color[HTML]{FFFFFF} \textbf{92.4}} & \cellcolor[HTML]{EFEFEF}3.6 & 0.0 \\
                                & \multicolumn{1}{l|}{4}                         & 0.0                                                          & 0.0                         & 0.0 & 0.0 &  & 0.0 & 0.0                                                          & 0.0                         & 0.0 &  & 4.8                                                          & 0.2                         & 0.0 & 0.0 &  & 0.0 & \cellcolor[HTML]{EFEFEF}3.8                                  & 0.0                         & 0.0 \\
\multicolumn{1}{l}{}            & \multicolumn{1}{r|}{5}                         & 0.0                                                          & 0.0                         & 0.0 & 0.0 &  & 0.0 & 0.0                                                          & 0.0                         & 0.0 &  & 0.0                                                          & 0.0                         & 0.0 & 0.0 &  & 0.0 & 0.2                                                          & 0.0                         & 0.0
\end{tabular}
\end{table}

\end{landscape}

\section{Application to patients with COPD} \label{secc_application}

Chronic obstructive pulmonary disease (COPD) is considered a complex, heterogeneous, and multisystemic disease \citep{vanfleteren2016management}. Therefore, tools are needed to assess all aspects of the disease in a comprehensive and integrated approach. Hospital admissions and mortality are two of the key outcomes of COPD.  To date, several models have been developed to predict adverse events in COPD patients \citep{esteban2008predictors,esteban2022changes,agusti2013characteristics}. In this context, accurate assessment of exercise capacity (six-minute walk test)  is considered an important variable due to the close relationship it has with patients’ health and the development of adverse events \citep{spruit2012predicting}. In fact, the predictor variable six-minute walk test has been widely used in the creation of prognosis scales as BODE index \citep{celli2004body}, defined on the basis of four variables: Body Mass Index (BMI), Obstruction (forced expiratory volume in one second, FEV1), Dyspnea (MMRC scale), and Exercise (6MWT).

For illustrative purposes, in this paper, we show the usefulness and interest of the proposed methodology by applying it to a longitudinal database of patients with stable COPD, in order to jointly categorize the variables FEV1 and 6MWT in a multiple Poisson model. Specifically, in this study patients being treated for COPD at five  outpatient respiratory clinics affiliated with the Galdakao Hospital in Biscay between January 2003 and January 2004 were recruited. Patients were consecutively included in the study if they had been diagnosed with COPD for at least six months and had been receiving medical care at one of the hospital's respiratory outpatient facilities for at least six months. Their COPD had to be stable for six weeks before enrollment. Patients were assessed at baseline and the end of the first, second, and fifth years, and were followed up for ten years. 

\begin{figure}
	\centering
	\includegraphics[width=0.6\textheight]{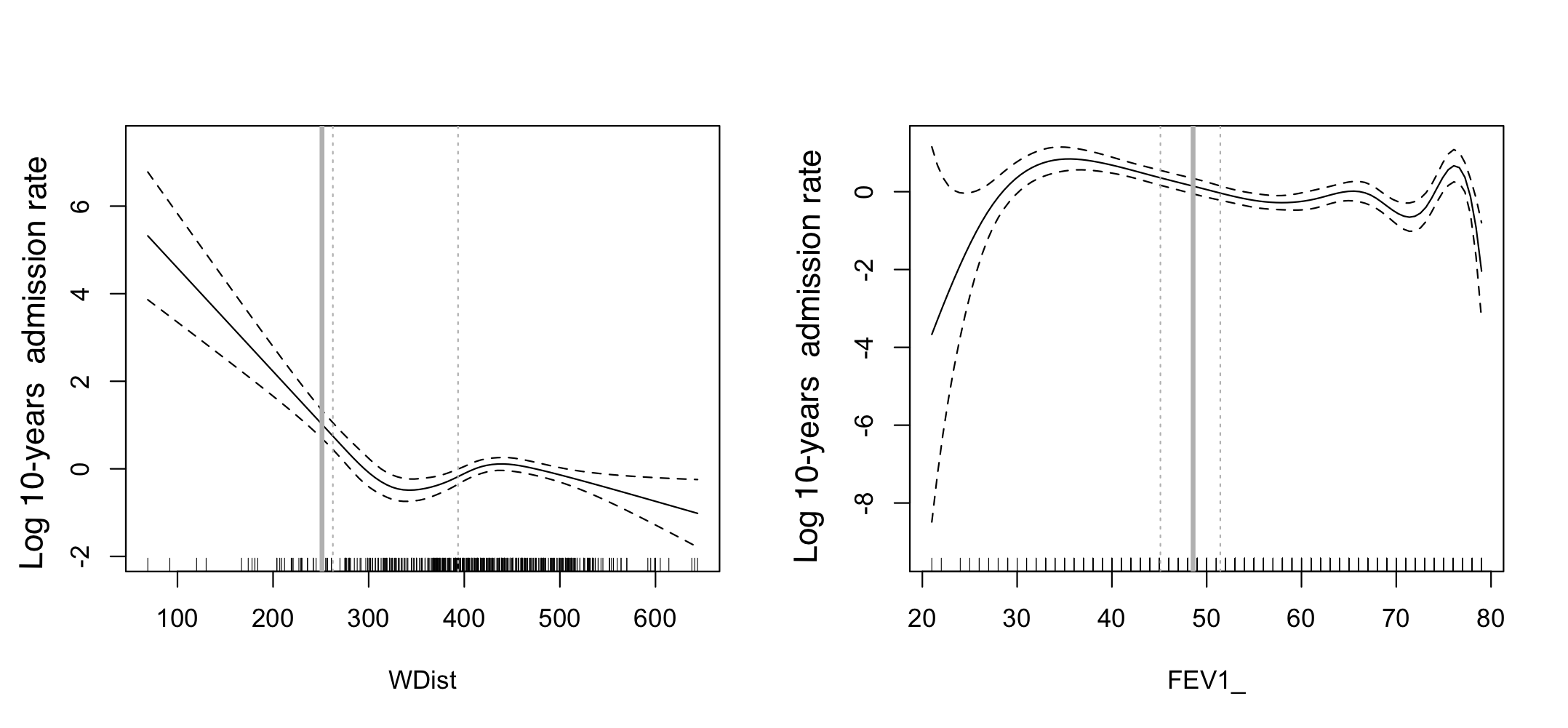}
	\caption{Estimated relationship between the predictor variables 6MWT and FEV1 and the log scale of 10-year admission rate in the multivariate Poisson GAM model. The solid line represents the location for $k=1$ cut-off point (selected number of cut-off points), while the dotted line represents the location for $k=2$ cut-off points.}
	\label{fig_copd}
\end{figure}

We considered the  10-year admission rate per number of years of follow-up as the response variable, for which a Poisson-GAM model was considered. The predictor variables 6MWT and FEV1, measured at the beginning of the study, were adjusted by means of a smooth function estimated using the adaptive fast estimation method  using the \texttt{sop} function of the \texttt{R} package \texttt{SOP},  in the presence of a set of statistically significant baseline covariates such as BMI, Charlson index, age, quadriceps or shoulder strength and the number of previous hospital admissions. The estimated relationship between the predictor variables 6MWT and FEV1 and the log scale of the 10-year admission rate in the multivariate Poisson GAM model can be observed in Figure \ref{fig_copd}.  We got that the best categorization was to consider a unique cut-off point for both 6MWT and FEV1 predictor variables, being the estimated cut-off points, 251 meters and 48.6 expiratory volume in percentage, respectively. These cut-off points are in line with the ones used previously in the literature \citep{celli2004body, vogelmeier2017global}.

The application of the proposed methodology to data from COPD patients with 10-year follow-up shows the usefulness and clinical practicality of the methodology, obtaining clinically validated results. Furthermore, this methodology allows us, unlike other existing categorization proposals, to obtain cut-off points simultaneously for more than one variable at the same time, and in the context of a Poisson model for rates.




\section{Discussion} \label{discussion}
	
Although different methods have been proposed to select the optimal location of the cut-off points to categorize (or dichotomize) a continuous predictor variable \citep{hin1999, tsuruta2006, barrio2013, Barrio2017, Barrio2017a} they are all developed to be applied in a particular regression context. To the best of our knowledge, up to now, no proposal has been made for the selection of the location nor the number of cut-off points in a global context of a GAM or GLM model for any response variable and which allows cut-off points to be obtained depending on the interaction of the variable of interest with another indicator variable. Therefore, in this paper, we have presented a new methodological proposal that allows us to respond to this need and which turns out to be computationally very efficient.  Different simulation studies have shown that this new proposal performs satisfactorily in regard to the location of the cut-off points and the selection of the number of categories, at least when we have a sufficient sample size to create the required number of categories. 

The proposed method has three main advantages. On the one hand, several variables can be categorized at the same time; secondly, it can be used for any response variable regardless of its distribution; and finally, the methodological idea can be easily extended to the context in which one wants to categorize a surface instead of a curve. 

With respect to the location of the cut-off points, we have seen that this method has a similar performance to previous proposals in the context of logistic regression (results not shown). However, the comparison with respect to the number of cut-off points is not trivial. The hypothesis testing proposed by \citet{Barrio2020} is computationally intractable when the objective is to categorize more than one variable at the same time. Therefore, we have proposed modified BIC criteria, similar to a variable selection process criteria, but penalizing for the number of cut-off points that are estimated. Computationally the BIC criterion is much faster and allows us to compare a large number of models (combinations of the number of categories) at the same time. However, unlike a hypothesis test, we cannot speak of type I errors and power rates. Based on the results obtained in the simulation study, we can say that the pseudo-BIC criterion is a conservative criterion, particularly when the available sample size is small. However, and despite the fact that this implies selecting fewer cut-off points than a priori necessary, we consider that it provides us with more parsimonious models.   

The methodology presented in this paper was applied to a COPD patient's data set where we categorized two continuous covariates, 6MWT and FEV1, simultaneously in a Poisson regression model. These two variables have been widely used in predictive models to predict the evolution of COPD patients \citep{agarwala2020six}. An example of this is 'indexes such as BODE \citep{celli2004body}, where these variables are categorized in the creation of the score. The search for cut-off points for variables such as 6MWT is of current interest, but it is common to search for a single cut-off point \citep{spruit2012predicting}.  The methodological approach allows us to find cut-off points for several variables simultaneously, obtaining cut-off points in a robust way. To the best of our knowledge, cut-off points for these variables have not been sought jointly in previous studies and neither in the specific context of a Poisson model. In addition, the results obtained are in line with the cut-off points used so far in the literature.

In summary, we consider that the existence of a valid statistical methodology to categorize continuous variables is of great importance to support researchers in other areas. In many cases, continuous variables are categorized without a clear criterion (equidistant intervals or quantiles are often used). The methodology proposed in this paper ensures that the cut-off points as well as the categories obtained offer a model whose results in terms of goodness of fit and discrimination ability are very close to those obtained with the GAM model for the continuous variable, and also offer a clear advantage in terms of practical application and interpretability.


\section*{Acknowledgements}
This work was financially supported in part by grants from the Departamento de Educación, Política Lingüística y Cultura del Gobierno Vasco IT1456-22 and by the Ministry of Science and Innovation through BCAM Severo Ochoa accreditation CEX2021-001142-S / MICIN / AEI /10.13039/501100011033 and through project PID2020-115882RB-I00 / AEI / 10.13039/501100011033 funded by Agencia Estatal de Investigación and acronym “S3M1P4R" and also by the Basque Government through the BERC 2022-2025 program and the BMTF ‘‘Mathematical Modeling Applied to Health’’ Project. The work of IB was supported by the Doctoral Research Staff Mobility Program of the Department of Education of the Basque Government.

\bibliographystyle{apalike}
\bibliography{biblio_paper}

\appendix

\end{document}